\documentstyle[12pt]{article}

\begin{document}


\def\a{\alpha}
\def\b{\beta}
\def\c{\varepsilon}
\def\d{\delta}
\def\e{\epsilon}
\def\f{\phi}
\def\g{\gamma}
\def\h{\theta}
\def\k{\kappa}
\def\l{\lambda}
\def\m{\mu}
\def\n{\nu}
\def\p{\psi}
\def\q{\partial}
\def\r{\rho}
\def\s{\sigma}
\def\t{\tau}
\def\u{\upsilon}
\def\v{\varphi}
\def\w{\omega}
\def\x{\xi}
\def\y{\eta}
\def\z{\zeta}
\def\D{\Delta}
\def\G{\Gamma}
\def\L{\Lambda}
\def\F{\Phi}
\def\P{\Psi}
\def\S{\Sigma}

\def\o{\over}

\def\IJMP{Int.~J.~Mod.~Phys. }
\def\MPL{Mod.~Phys.~Lett. }
\def\NP{Nucl.~Phys. }
\def\PL{Phys.~Lett. }
\def\PR{Phys.~Rev. }
\def\PRL{Phys.~Rev.~Lett. }
\def\PTP{Prog.~Theor.~Phys. }
\def\ZP{Z.~Phys. }

\def\beq{\begin{equation}}
\def\eeq{\end{equation}}


\title{
  \begin{flushright}
    \large UT-773
  \end{flushright}
  \vspace{5ex}
  Strongly Coupled Messenger Gauge Theory}
\author{Izawa K.-I. \\
  \\  Department of Physics, University of Tokyo \\
  Bunkyo-ku, Tokyo 113, Japan}
\date{April, 1997}
\maketitle

\begin{abstract}
We consider gauge-mediated models of supersymmetry (SUSY) breaking
with strongly coupled nonabelian messenger gauge theory.
After integrating out the SUSY-breaking and messenger sectors,
we directly see that the effective K{\" a}hler potential
transmits the SUSY breaking to the standard model sector.
\end{abstract}

\newpage

\section{Introduction}

Supersymmetry (SUSY), if present in nature, is broken at low energy
so that the observed particles do not
have their superpartners with the same masses.
Spontaneous SUSY breaking is naturally realized through
dynamics of nonabelian gauge theory.
Transmission of breaking
from dynamical SUSY breaking to the standard model sectors
is achieved by means of various interactions
such as gravity and gauge interactions
\cite{Din}.

In gauge-mediated scenario of SUSY breaking,
weakly coupled messenger gauge theory is conventionally used
\cite{Nel}
to mediate the breaking naturally to the standard model sector.
In this paper, we take strongly coupled nonabelian theory as messenger gauge
theory and see its contents by means of effective theory analysis.

\section{The SUSY-Breaking Sector}

Let us first consider the SUSY-breaking sector.
We take a SUSY SU(2)$_S$ gauge theory with four doublet chiral
superfields $Q_i$ and six singlet ones $Z^{ij} = -Z^{ji}$.
Here $i$ and $j$ denote the flavor indices ($i, j = 1, \cdots, 4$).
Without a superpotential, this model has a flavor SU(4)$_F$ symmetry.

The tree-level superpotential of the model \cite{Iza} is given by
\begin{equation}
  W_{tree} = \lambda_{ij}^{kl} Z^{ij} Q_k Q_l,
\end{equation}
where $\lambda_{ij}^{kl}$ denote generic coupling constants with
$\lambda_{ij}^{kl} = -\lambda_{ji}^{kl} = -\lambda_{ij}^{lk}$.
SUSY remains unbroken perturbatively in this model.

The effective superpotential of the model
may be written in terms of
gauge-invariant low-energy degrees of freedom
\begin{equation}
  V_{ij} = -V_{ji} \sim Q_i Q_j
\end{equation}
as follows:
\begin{equation}
  W_{eff} = X({\rm Pf} \, V_{ij} - \Lambda_S^4) + \lambda_{ij}^{kl}
  Z^{ij} V_{kl},
\end{equation}
where $X$ is an additional chiral superfield, ${\rm Pf} \, V_{ij}$
denotes the Pfaffian of the antisymmetric matrix $V_{ij}$, and
$\Lambda_S$ is a dynamical scale of the SU(2)$_S$ gauge interaction.
This effective superpotential yields conditions for SUSY vacua
\begin{equation}
  {\rm Pf} \, V_{ij} = \Lambda_S^4, \quad \lambda_{ij}^{kl} V_{kl} = 0,
\end{equation}
which are incompatible as long as $\Lambda_S \neq 0$.
Therefore we conclude that SUSY is dynamically broken in this model
\cite{Iza}.

Let us impose a flavor SP(4)$_F$ ($\subset {\rm SU}(4)_F$)
symmetry on the above model to make
our analysis simpler.
Then the effective superpotential can be written as
\begin{equation}
  \label{dsbPot}
  W_{eff} = X(V^2 + V_a V_a - \Lambda_S^4) + \lambda_Z Z V + \lambda Z^a
  V_a,
\end{equation}
where $V$ and $Z$ are singlets and $V_a$ and $Z^a$ are
five-dimensional representations of SP(4)$_F$, respectively, in
$V_{ij}$ and $Z^{ij}$, which constitute six-dimensional
representations of SU(4)$_F$.
Here $a = 1, \cdots, 5$ and $\lambda_Z$ and $\lambda$ denote coupling
constants which are taken to be positive.

When the coupling $\lambda_Z$ is small, the effective superpotential
$W_{eff}$ implies that we obtain the following vacuum expectation
values:
\begin{equation}
  \label{vevs}
  \langle V \rangle \simeq \Lambda_S^2, \quad \langle V_a \rangle \simeq
  0.
\end{equation}
Then the low-energy effective superpotential may be approximated by
\begin{equation}
  \label{effPot}
  W_{eff} \simeq \lambda_Z \Lambda_S^2 Z.
\end{equation}

On the other hand, the effective K\"ahler potential is expected to
take a form
\begin{equation}
  \label{kahler}
  K = |Z|^2 - \frac{\eta}{4 \Lambda_S^2} \lambda_Z^4 |Z|^4 + \cdots,
\end{equation}
where $\eta$ is a real constant of order one.

Then the effective potential of the scalar $Z$ (with the same notation
as the superfield) is given by
\begin{equation}
  V_{eff} \simeq \lambda_Z^2 \Lambda_S^4 (1 + \frac{\eta}{\Lambda_S^2}
  \lambda_Z^4 |Z|^2 + \cdots).
\end{equation}
When $\eta > 0$, this leads to $\langle Z \rangle = 0$.
Otherwise, we suspect that $\langle Z \rangle$ is of order
$\lambda_Z^{-1} \Lambda_S$
\cite{Iza,Hot,Shi}.
We henceforth assume $\eta > 0$ for definiteness
and consider the situation with $\langle Z \rangle = 0$.

\section{The Messenger Sector}

The messenger sector consists of chiral superfields
$Q'_I, Y, d, \bar{d}, l, \bar{l}$,
which are all singlets under the SUSY-breaking SU(2)$_S$.
Here $I$ denotes a flavor index: $I = 1, 2$.
$Q'_I$ and $Y$ are also singlets
under the standard-model gauge group,
while $d, \bar{l}$ and $\bar{d}, l$ are assumed to
transform as the down quark, the anti-lepton doublet and their
antiparticles, respectively.

We further introduce a messenger gauge group SU(2)$_M$
under which $Q'_I$ constitute a pair of doublets.
The other superfields are all singlets under the messenger SU(2)$_M$.

The interactions in the messenger sector are
given by a superpotential
\begin{equation}
 \label{intPot}
 W_{mess} = \lambda_Y Q'_1 Q'_2 Y
          - \frac{f}{3} Y^3 + (k_1 \bar{d} d + k_2 \bar{l} l) Y,
\end{equation}
where
the couplings $\lambda_Y$, $f$, $k_1$ and $k_2$ are taken to be positive.
In the following, we set $k \equiv k_1 = k_2$, which
approximately holds in SUSY-GUT's.

The effective superpotential of the messenger sector
is given by
\begin{equation}
 \label{MEFF}
 W_{eff} = 2 (\l_Y^{1 \o 5} \L_M)^{5 \o 2} Y^{1 \o 2}
         - \frac{f}{3} Y^3 + k(\bar{d} d + \bar{l} l)Y,
\end{equation}
where $\L_M$ denotes a dynamical scale of the SU(2)$_M$ messenger
gauge interaction.
This superpotential leads to two SUSY vacua:
$\langle Y \rangle = f^{-{2 \o 5}} \l_Y^{1 \o 5} \L_M,
\langle \bar{d} \rangle = \langle d \rangle = \langle \bar{l} \rangle
= \langle l \rangle = 0$ and
$\langle Y \rangle = 0,
\langle \bar{d} d + \bar{l} l \rangle \rightarrow \infty$.
When the SUSY-breaking is transmitted to the messenger sector
(in the next section),
these vacua may be lifted to yield SUSY-breaking vacua.%
\footnote{These vacua might be local minima, as is the case for the
weakly coupled messenger gauge theory \cite{Das}.}
We restrict ourselves to the vacuum corresponding to the former one
in the following analysis of the SUSY-breaking case
since, otherwise, the standard-model gauge
group would be broken by the vacuum expectation values
of the messenger quarks and leptons.

\section{The Combined System}
\label{mu}

In order to combine the SUSY-breaking and messenger sectors,
we introduce mediator chiral superfields $Q''_A$ which are doublets under both
the SUSY-breaking SU(2)$_S$ and messenger SU(2)$_M$ gauge groups.
Here $A$ denotes a flavor index: $A = 1, \cdots, N$.
The superpotential of our model for the mediators
is given by
\begin{equation}
 \label{REQ}
 W_{med} = m Q''_A Q''_A,
\end{equation}
where $m$ is a mass parameter.%
\footnote{Mass terms for SUSY-breaking transmission
were considered in Ref.\cite{Hot,Ran}.
We note that it is possible to consider models of SUSY-breaking
transmission with massive messenger
quarks and leptons which are also charged under the gauge symmetry
of dynamical SUSY breaking.}
This is a SUSY mass term similar to that for
the Higgs doublets in the SUSY standard model (so-called $\m$ term).
In this paper, we postulate that the scale $m$ comes out along with the
$\m$ term through some mechanism.%
\footnote{For example, we may rely on a similar mechanism to that
utilized in order to generate the SUSY-breaking scale in section 2.}

In the following, we assume
\begin{equation}
 \label{SCA}
 \L_S \simeq \L_M \simeq m \equiv \L
\end{equation}
for simplicity (see the final section).
Note that
the mediators $Q''_\a$ may be integrated out since they are massive.

In view of Eqs.(\ref{effPot}) and (\ref{MEFF}),
the full effective superpotential of the SUSY-breaking and messenger sectors
is obtained as
\begin{equation}
 \label{messPot}
 W_{eff} \simeq \lambda_Z \Lambda^2 Z
         + 2 (\lambda_Y^{1 \o 5} \Lambda)^{5 \o 2} Y^{1 \o 2}
         - \frac{f}{3} Y^3 + k ( \bar{d} d + \bar{l} l) Y,
\end{equation}
since the superpotential does not suffer from renormalization
due to the massive mediators.

On the other hand,
the effective K\"ahler potential of $Z$ and $Y$ is expected to have the
following form:
\begin{equation}
 \label{Kahler}
 K = |Z|^2 + |Y|^2
   - \frac{\eta}{4 \Lambda^2} \lambda_Z^4 |Z|^4
   - \frac{\delta}{\Lambda^2} \lambda_Z^2 \l_Y^2 |Z|^2 |Y|^2
   + \cdots,
\end{equation}
where $\delta$ is a real constant of order one,
which we assume to be positive for definiteness.
The mixing term of the form $|Z|^2 |Y|^2$ stems from the interaction
of the massive mediators.

By means of Eqs.(\ref{messPot}) and (\ref{Kahler}),
we obtain an effective potential, whose relevant part for the present analysis
is given by
\begin{equation}
 \label{fullPot}
 V_{eff} \simeq \lambda_Z^2 \Lambda^4 (1 + {\d \o \L^2} \l_Z^2 \l_Y^2 |Y|^2)
         + |{(\lambda_Y^{1 \o 5} \Lambda)^{5 \o 2} \o Y^{1 \o 2}} - f Y^2|^2.
\end{equation}
This potential yields a vacuum
\begin{equation}
  \langle F_Y \rangle
  \simeq \langle {(\lambda_Y^{1 \o 5} \Lambda)^{5 \o 2} \o Y^{1 \o 2}} - f Y^2
  \rangle
  \simeq \delta {2 \o 5}
  f^{-1} \lambda_Z^4 \lambda_Y^2 \Lambda^2, \quad
  \langle Y \rangle \simeq f^{-{2 \o 5}} \lambda_Y^{1 \o 5} \Lambda,
\end{equation}
where we have taken into account only the leading terms in the
coupling $\lambda_Y$.

\section{Conclusion}
\label{conclusion}

The standard-model gauginos obtain their masses from radiative
corrections through loops of the messenger quarks and leptons
\cite{Nel}:
\begin{equation}
  \label{gauginomass}
  m_{\tilde{g_i}}
  \simeq \frac{\alpha_i}{4 \pi} \frac{\langle F_Y \rangle}{\langle Y
    \rangle}
  \simeq \frac{\alpha_i}{4 \pi}
  \delta \frac{2}{5}
  f^{-{3 \o 5}} \lambda_Z^4 \lambda_Y^{9 \o 5} \Lambda,
\end{equation}
where $\alpha_i = g_i^2/4\pi$ denote the standard-model gauge
couplings.
In terms of the parameters
$\lambda_Z \simeq \l_Y \simeq f \simeq \d \simeq 1$, for
instance, we get $m_{gluino} \sim 10^3$ GeV for $\L \sim 10^5$ GeV.

Finally we comment on the dynamical scales $\L_S$ and $\L_M$.
Let $\L_{S0}$ and $\L_{M0}$ be the dynamical scales
determined by the SU(2)$_S$ and SU(2)$_M$
gauge couplings well above the mass scale $m$.
Then the dynamical scales at high and low energies are related by
\cite{Fin}
\beq
 \L_S \sim (\L_{S0}^{4-N} m^N)^{1 \o 4}, \quad
 \L_M \sim (\L_{M0}^{5-N} m^N)^{1 \o 5},
\eeq
which implies that $\L_S$ and $\L_M$ are expected to be close to the
mass scale $m$ for appropriate values of $N$.
Namely, the SUSY-breaking scale may be naturally
controlled by the SUSY-invariant mass in the present scheme,
which justifies the simple assumption (\ref{SCA}) for the analysis.

\section*{Acknowledgement}

The author would like to thank T.~Yanagida for valuable discussions.


\newpage

\begin{thebibliography}{99}

\bibitem{Din}
  For a review, M.~Dine, hep-ph/9612389.

\bibitem{Nel}
  M.~Dine and A.E.~Nelson, Phys.~Rev. {\bf D48} (1993) 1277; \\
  M.~Dine, A.E.~Nelson, and Y.~Shirman, Phys.~Rev. {\bf D51} (1995) 1362; \\
  M.~Dine, A.E.~Nelson, Y.~Nir, and Y.~Shirman,
  Phys.~Rev. {\bf D53} (1996) 2658.

\bibitem{Iza}
  Izawa~K.-I. and T.~Yanagida, Prog.~Theor.~Phys. {\bf 95} (1996) 829; \\
  K.~Intriligator and S.~Thomas, \NP {\bf B473} (1996) 121.

\bibitem{Hot}
  T.~Hotta, Izawa~K.-I., and T.~Yanagida, \PR {\bf D55} (1997) 415.

\bibitem{Shi}
  See also Y.~Shirman, \PL {\bf B389} (1996) 287.

\bibitem{Das}
  I.~Dasgupta, B.A.~Dobrescu, and L.~Randall, \NP {\bf B483} (1997) 95.

\bibitem{Ran}
  L.~Randall, hep-ph/9612426.

\bibitem{Fin}
  D.~Finnell and P.~Pouliot, \NP {\bf B453} (1995) 225, and references therein.

\end{thebibliography}
\end{document}